# Analytic study on cold field electron emission from metallic nanowall array[*]

Xizhou Qin, Weiliang Wang and Zhibing Li[a)]

*State Key Laboratory of Optoelectronic Materials and Technologies,*

*School of Physics and Engineering, Sun Yat-Sen University, Guangzhou 510275, China*

The cold field electron emission from metallic nanowall array is investigated theoretically. Via conformal mapping method, analytic formulas of tunneling barrier, edge field enhancement factor, transmission coefficient, and area emission current density are derived in the semi-classical approximation. The influence of field enhancement effects and screening effects on cold field emission are discussed and the optimal spacing distance between nanowalls for the largest emission current density is found to be a linear function of the height of nanowalls and depends on the applied field.

**Keywords:** metallic nanowall array, cold field electron emission, field enhancement effects, screening effects

**PACS:** 79.70.+q, 68.37.Vj

## 1. Introduction

Cold field electron emission (CFE) as an important vacuum electron source has applications in flat-panel display, electronic holography, e-beam lithography and so on. Emitters of various nanostructures, especially nanotubes and nanowires, have been extensively studied. The major merit of nano-emitters is that an applied electric field can be greatly enhanced around the nano-scaled tips or edges. Since the experimental realization of free-standing graphene [1], this two-dimensional (2D) atomic crystal has aroused great experimental and theoretical interest. Several groups have demonstrated that graphene does show promising CFE properties such as a low emission threshold field and large emission current density. [2-8]

Recent studies show that the CFE from the 2D nanostructures would have a current-field characteristic different from that of the conventional Fowler–Nordheim

[*] The Project was supported by the National Basic Research Program of China (Grant No. 2007CB935500) and the National Natural Science Foundation of China (Grant No. 11104358).
[a)] Corresponding author. E-mail: stslzb@mail.sysu.edu.cn

(FN) law which was derived for planar emitters [9]. It is obvious because both electron supply function and edge electric field of the 2D nanostructures are dramatically different from those of the three-dimensional systems. The long edge of 2D emitters with atomic thickness may lead to a unique line-electron source that would be useful in electronic holography and parallel e-beam lithography.

The present paper is interested in the aligned array of 2D metallic nanowalls that is defined as an array of parallel blade-like conducting structures mounted vertically on a planar cathode. It can be anticipated that the spacing distance between emitters and the height of the nanowall are two crucial parameters for the total CFE current density. A narrow spacing will reduce the edge electric field due to the strong screening effect, and then reduce the CFE current of each nanowall. On the other hand, larger distance between the emitters means less number of nanowalls in a given area of cathode. Obviously there is an optimistic spacing that gives the largest emission current density. The aim of this paper is to find out this optimistic spacing.

Section 2 describes our model and introduces the conformal mapping method, which is usually used to find the solution of the Laplace equation with 2D boundary condition [10-15]. Section 3 is divided into three subsections. Subsection 3.1 solves the 2D Laplace equation and gives the explicit expression of the vacuum electric potential for the forward emission; subsection 3.2 derives the enhancement factor and discusses the screening effects; subsection 3.3 gives the transmission coefficient and the forward emission current density via the semi-classical (JWKB) approximation. The angle-dependent transmission coefficient is also calculated to support that the forward emission is dominant. The main results will be summarized in Section 4.

## 2. Model and method

Let us consider an array of nanowalls mounted on a planar cathode vertically. The sketch and cross section of the set up are shown in Fig. 1, where the bottom and the top planes are the cathode and anode respectively. The nanowalls have the same width $2r$, the same height $h$, and infinite length in the $z$ direction. The spacing distance between two neighboring nanowalls (NWSD) is denoted by $2d$. To simplify the analysis, we assume that nanowalls are all metallic. The cathode and nanowalls are grounded and their electric potential is pinned at zero. To make the problem analytically tractable, we assume the anode is remote from the cathode ($X_a \to \infty$), and the macroscopic applied electric field $F_M = V_{app}/X_a$ has a fixed value while $V_{app} \to \infty$, with $V_{app}$ the voltage applied to the anode.

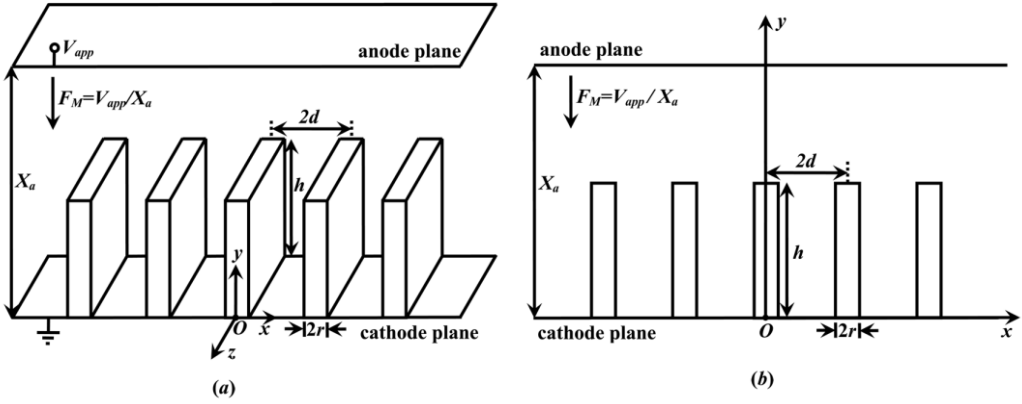

**Fig. 1.** Setup for field emission from metallic nanowall array. All nanowalls are vertically mounted on the cathode (bottom plane). (a) Three-dimensional view; (b) projection onto the *x-y* plane.

Electric potential in the vacuum gap (VGEP) between the cathode and anode is denoted by $V(x, y)$, which satisfies the 2D Laplace equation and the boundary conditions that $V$ is zero on the surfaces of cathode and nanowalls, and the partial derivative $-\partial V/\partial y$ is $F_M$ when $y \to \infty$. The complex coordinate $z = x + iy$ will be used to specify points of the *x-y* plane.

Two successive conformal transformations (CTs) are introduced to find the VGEP. As shown in Fig. 2, the CT from (b) to (a) maps a virtual space ($\zeta'$) into the physical space ($z$), and the CT from (c) to (b) maps the target space ($\zeta$), where the potential is easy to obtain, into the virtual space ($\zeta'$). The representative points $W_0$, $W_{\sigma'}$, $W_1$, and $W_{\sigma''}$ in the physical space correspond to points $0$, $\sigma'$, $1$ and $\sigma''$ on the real axis of the virtual space. The parameters $\sigma'$ and $\sigma''$ will be specified in the following (Eqs. (3) and (4)). We have required $0 < \sigma' < 1 < \sigma''$ for the latter convenience.

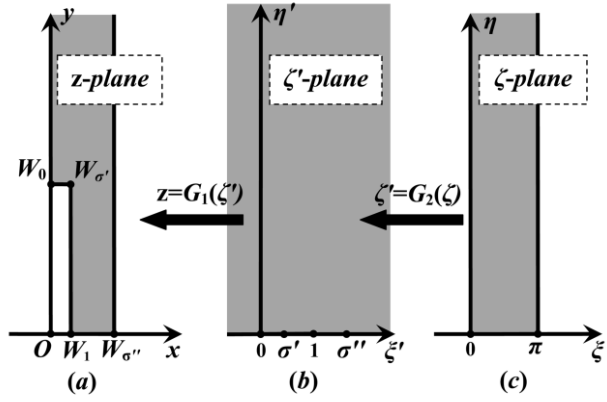

**Fig. 2.** (a) The physical space ($z$); (b) the virtual space ($\zeta'$); and (c) the target space ($\zeta$). These spaces are connected by two conformal transformations $G_1$ and $G_2$.

Denote the CT from Fig. 2(b) to Fig. 2(a) by $z = G_1(\zeta')$. It is given by Schwarz–Christoffel formula [16],

$$G_1(\zeta') = i\left[\frac{r}{B(0,\sigma')}\int_0^{\zeta'}\omega^{-1/2}(\omega-\sigma')^{1/2}(\omega-1)^{-1/2}(\omega-\sigma'')^{-1/2}d\omega + h\right], \quad (1)$$

where $B(t_0,t_1)$ is a real function of $t_0$ and $t_1$ defined by

$$B(t_0,t_1) = \int_{t_0}^{t_1}\left|t^{-1/2}(t-\sigma')^{1/2}(t-1)^{-1/2}(t-\sigma'')^{-1/2}\right|dt. \quad (2)$$

This CT maps the representative points $0$, $\sigma'$, $1$ and $\sigma''$ in the virtual space to the points $W_0$, $W_{\sigma'}$, $W_1$, and $W_{\sigma''}$ in the physical space, respectively. The requirements of $W_0 = ih$, $W_{\sigma'} = r+ih$, $W_1 = r$, and $W_{\sigma''} = d$ can be fulfilled if the parameters $\sigma'$ and $\sigma''$ satisfy the relations

$$\frac{B(0,\sigma')}{B(\sigma',1)} = \frac{r}{h}, \quad (3)$$

$$\frac{B(1,\sigma'')}{B(0,\sigma')} = \frac{d}{r} - 1. \quad (4)$$

Equations (3) and (4) specify the parameters $\sigma'$ and $\sigma''$. The values of $B(t_0,t_1)$ in Eqs. (3) and (4) can be reduced to complete elliptic integrals (Appendix A) that depend only on the parameters $\sigma'$ and $\sigma''$,

$$B(0,\sigma') = c(1-\sigma')[\Pi_1(\sigma',m) - K(m)], \quad (5)$$

$$B(\sigma',1) = c\sigma'\left[\Pi_1(1-\sigma',m') - K(m')\right], \tag{6}$$

$$B(1,\sigma'') = c(1-\sigma')\Pi_1(b,m), \tag{7}$$

where we have defined $b = (\sigma''-1)/(\sigma''-\sigma')$, $m = b\sigma'$, $m' = 1-m$, and $c = 2(\sigma''-\sigma')^{-1/2}$.

The second CT $\zeta' = G_2(\zeta)$ that maps the target space to the virtual space is

$$G_2(\zeta) = \sigma'' \sin^2(\zeta/2), \tag{8}$$

where $\zeta$ is the complex coordinate of the target space.

## 3. Results and discussion

### 3.1. Electric potential

The electric potential $V_{\text{target}}(\zeta)$ in the target space (Fig. 2(c)) is a uniform field. The solution reads,

$$V_{\text{target}}(\zeta) = F_M \cdot r/B(0,\sigma') \cdot \text{Im}(\zeta). \tag{9}$$

Denote the inversed function of $z = G_1(\zeta')$ and $\zeta' = G_2(\zeta)$ as $\zeta' = G_1^{-1}(z)$ and $\zeta = G_2^{-1}(\zeta')$, respectively. Then the electric potential in the physical space has a formal expression,

$$V(x,y) = V_{\text{target}}(G_2^{-1}(G_1^{-1}(x+iy))). \tag{10}$$

Because $G_1^{-1}(z)$ is very complicated, it is difficult to use Eq. (10) to calculate $V(x,y)$ directly even when the parameters $\sigma'$ and $\sigma''$ are known. But as will be shown below, one can make a good estimation for the emission current density via just knowing the potential along the $y$-axis as the forward emission is dominant.

The function $G_1(\zeta')$ can be expressed in elliptic integrals at the real axis of the virtual space. The point $z = 0+iy$ on the $y$-axis of the physical space for $y \geq h$ is corresponding to the point $\zeta' = \xi'+i0$ on the $\xi'$-axis of the virtual space for

$\xi' \leq 0$, while the latter is corresponding to the point $\zeta = 0 + i\eta$ on the $\eta$-axis of the target space for $\eta \geq 0$.

From Eqs. (1) and (8), one has,

$$y = h + (d-r)\frac{\sigma'}{1-\sigma'} \cdot \frac{\Pi(b', \varphi_{\xi'} | m')}{\Pi_1(b, m)}, \tag{11}$$

$$\xi' = -\sigma'' \sinh^2(\eta/2), \tag{12}$$

where we have defined $b' = \sigma''/(\sigma'' - \sigma')$, and $\varphi_{\xi'} = \arcsin(\sqrt{-\xi'/[b'(-\xi' + \sigma')]})$.

Denote the inverse function of $u = \Pi(b', \arcsin t | m')$ as $t = \Phi_{b',m'}(u)$. Then Eqs. (11) and (12) read

$$\eta = 2 \cdot \text{arcsinh}\left(\sqrt{\frac{\sigma'}{\sigma''} \frac{b' \Phi_{b',m'}^2(k(y-h))}{1 - b' \Phi_{b',m'}^2(k(y-h))}}\right), \tag{13}$$

where we have defined $k = (1-\sigma')\Pi_1(b,m)/(\sigma'(d-r))$.

With this and Eq. (9) and (10), it follows that,

$$V(0, y) = 2F_M \frac{r}{B(0, \sigma')} \text{arcsinh}\left(\sqrt{\frac{\sigma'}{\sigma''} \frac{b' \Phi_{b',m'}^2(k(y-h))}{1 - b' \Phi_{b',m'}^2(k(y-h))}}\right). \tag{14}$$

This is the expression for electric potential along the $y$-axis. Figure 3 shows the electric potentials along the $y$-axis with various $d$. The inset shows that the electric potential is approximately linear with $(y-h)^{1/2}$ when $0.3 \text{ nm} < y-h < 100.0 \text{ nm}$. The $d$-dependence shows the screening effect.

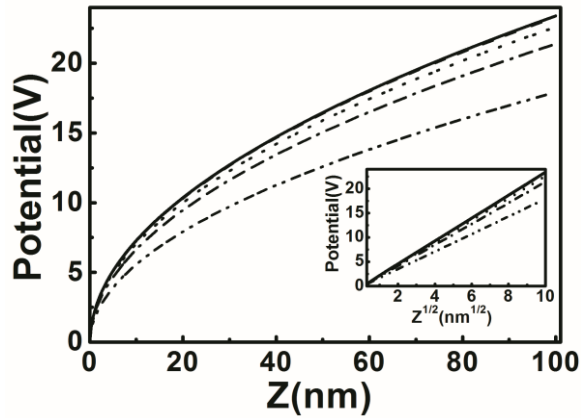

**Fig. 3.** Electric potentials versus $Z = y - h$ with $d = 3.0$ μm (dash-dot-dot), $6.0$ μm (dash-dot), $10.0$ μm (dot), $30.0$ μm (dash), and infinite NWSD (solid). The $Z$-axis has origin at the center of the edge of a nanowall of $3.0$ μm in height and $0.16$ nm in width. The applied field is $30.0$ V/μm. The inset is electric potentials versus $Z^{1/2}$. The curves with $d = 30.0$ μm and infinite NWSD coincide with each other.

## 3.2. Field enhancement factor

Note that $\Phi_{b',m'}(0) = 0$. Hence the r.h.s. of Eq.(14) is $0$ at $y = h$, as the boundary conditions required. Using $d\Phi_{b',m'}(u)/du = dt/d\Pi(b', \arcsin t | m')$ at $y = h$, the electric field at the middle of the nanowall top can be obtained from Eq. (14),

$$F(0,h) = -\frac{dV(0,y)}{dy}\bigg|_{y=h} = -F_M \sqrt{\frac{1}{\sigma'}}. \tag{15}$$

Thus it is straight forward to obtain the field enhancement factor, $\beta = F(0,h)/(-F_M)$, at the middle of the nanowall top, resulting

$$\beta = \sqrt{\frac{1}{\sigma'}}. \tag{16}$$

In the limit of $d \to \infty$, Eq. (4) imply $\sigma'' \to \infty$. Substitute the infinite $\sigma''$ limits of $B(0,\sigma')$ and $B(\sigma',1)$ into Eq. (3),

$$\frac{1-\sigma'}{\sigma'} \frac{\Pi_1(\sigma',\sigma') - K(\sigma')}{\Pi_1(1-\sigma',1-\sigma') - K(1-\sigma')} = \frac{r}{h}. \tag{17}$$

Using the identity $\Pi_1(n,n) = E(n)/(1-n)$, Eq.(17) can be reduced to

$$\frac{r}{h} = \frac{\rho E_s(\rho)}{\sqrt{1-\rho^2} E_s(\sqrt{1-\rho^2})}, \tag{18}$$

where we have defined $\rho = \sqrt{\sigma'}$ and adopted the notation (Appendix A) $E_s(\rho) = E(\arcsin(\rho) | \rho^{-2})$.

Figure 4 shows that the field enhancement factor increases with the NWSD, and tends to a constant when NWSD tends to infinity. Actually (18) is the equation (5.2) of Ref. [9]. We see, as expected, the field enhancement of a single nanowall is recovered in large NWSD.

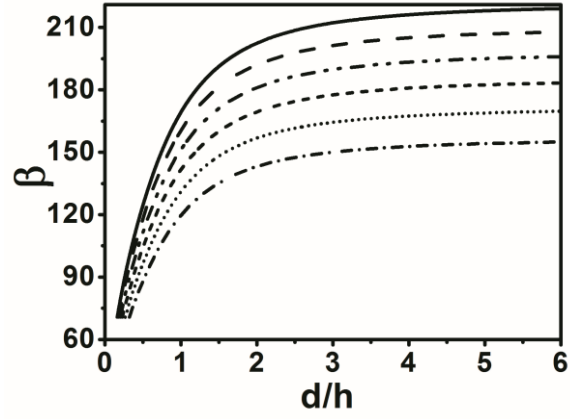

**Fig. 4.** Field enhancement factor $\beta$ versus $d/h$ with $h = 2.5\ \mu m$ (dash-dot), $3.0\ \mu m$ (dot), $3.5\ \mu m$ (short-dash), $4.0\ \mu m$ (dash-dot-dot), $4.5\ \mu m$ (dash), and $5.0\ \mu m$ (solid). The nanowall width is $0.16\ nm$ and each curve is obtained with a fixed $h$ and various NWSDs.

### 3.3. Transmission coefficient and current density

The transmission coefficient of the electron in the Fermi level $E_F$ can be estimated by the JWKB approximation,

$$D = \exp\left[-g_e \int_0^\Delta \sqrt{U(Z) - E_F}\, dZ\right], \qquad (19)$$

where $U(Z)$ is the tunneling barrier with the interval $(0, \Delta)$ as the classical forbidden region where $U(Z) - E_F > 0$, $\Delta$ the potential barrier width (BW), $g_e = 4\pi(2m_e)^{1/2}/h_P$ the JWKB constant for electron, $h_P$ the Planck constant, and $m_e$ the mass of an electron.

The BW can be calculated by Eq. (10). The angular dependence of BW (Fig. 5) shows that the forward emission is dominant. Along the forward direction ($y$-axis) a simple expression for BW $\Delta_y$ can be found (Appendix B)

$$\Delta_y = r \cdot \frac{\sigma'}{1-\sigma'} \cdot \frac{\Pi(b', \varphi_\Delta | m')}{\Pi_1(\sigma', m) - K(m)}, \qquad (20)$$

where $\phi$ is the work function of the nanowalls, $e_0$ the elementary positive charge and we have defined $\varphi_\Delta = \arcsin\sqrt{(\sigma'' - \sigma')/(\sigma'' + \sigma' \sinh^{-2}[\phi B(0, \sigma')/(2e_0 F_M r)])}$.

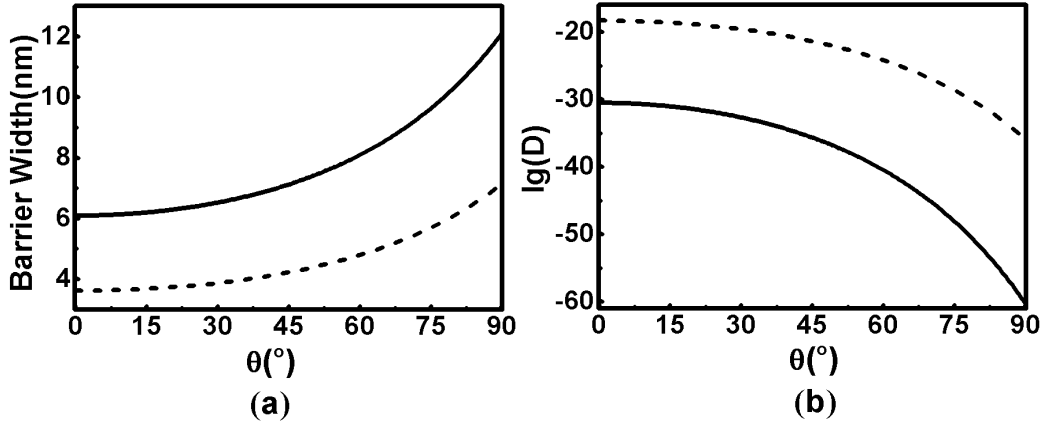

**Fig. 5.** (a) Potential barrier thickness $\Delta$ versus the angle $\theta$ between the emission direction and the $y$-axis with $d = 3.0$ μm (solid) and infinite NWSD (dotted). (b) Logarithmic plot of the transmission coefficient versus $\theta$. Each nanowall has width $0.16$ nm and height $3.0$ μm. The applied electric field is $F_M = 30.0$ V/μm and the work function is assumed to be $4.32$ eV.

From Eqs. (19) and (20), the transmission coefficient along the $y$-axis is (Appendix C)

$$D_y = \exp\left[-\frac{g_e \phi^{3/2} h_D}{e_0 F_M}\right], \tag{21}$$

where we have defined

$$h_D = \int_0^1 \sqrt{1-t} \sqrt{\frac{\sigma'' \sinh^2[\phi B(0,\sigma')t/(2e_0 F_M r)] + \sigma'}{\sigma'' \sinh^2[\phi B(0,\sigma')t/(2e_0 F_M r)] + 1}} dt. \tag{22}$$

The area emission current density at zero temperature is (Appendix C),

$$J_A^{nws}(F_M) = z^{nws} \cdot (2d)^{-1} \cdot \frac{1}{h_C^{3/2}} \cdot \frac{F_M^{3/2}}{\phi^{3/4}} \cdot \exp\left(-\frac{g_e h_D}{e_0} \frac{\phi^{3/2}}{F_M}\right), \tag{23}$$

where we have defined $z^{nws} = e_0^{5/2}/(2^{3/4} \pi h_P^{1/2} m_e^{1/4})$, and

$$h_C = \int_0^1 \frac{1}{\sqrt{1-t}} \sqrt{\frac{\sigma'' \sinh^2[\phi B(0,\sigma')t/(2e_0 F_M r)] + \sigma'}{\sigma'' \sinh^2[\phi B(0,\sigma')t/(2e_0 F_M r)] + 1}} dt, \tag{24}$$

Figure 6 are typical $J$-$d$ curves. Firstly, $J$ increases with small $d$ as the screening effect gets weaker, and then $J$ decrease with $d$ as the area density of nanowall decreases. The optimal $d$ is larger for higher nanowall. Yet more complicated DFT calculation should be adopted when $d$ is close to the thickness of the nanowall, as the interaction between adjacent nanowall becomes important [17].

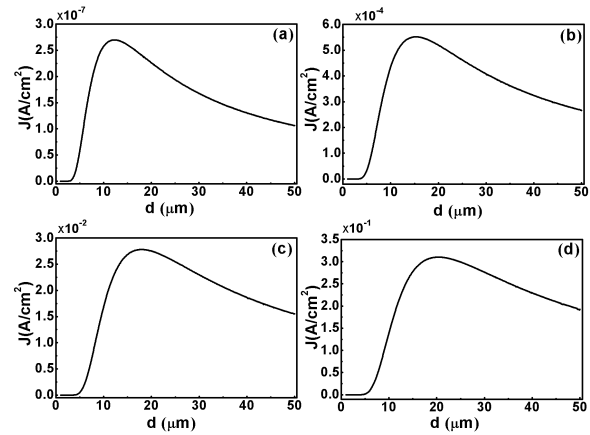

**Fig. 6.** Area current density $J_A^{nws}$ versus $d$ with (a) $h=2.0$ μm, (b) $h=3.0$ μm, (c) $h=4.0$ μm, (d) $h=5.0$ μm. The applied field $F_M$ is 50.0 V/μm, and the nanowall width is 0.16 nm.

Figure 7 shows the height-dependent optimal NWSD for different applied electric fields. The best fitting to the simulation (Fig. 7) gives,

$$d_{opt} \approx \frac{1}{-0.019+0.0028F_M} + \frac{h}{0.14+0.0055F_M}. \qquad (25)$$

Where $d_{opt}$ and $h$ are in unit of μm, $F_M$ in V/μm.

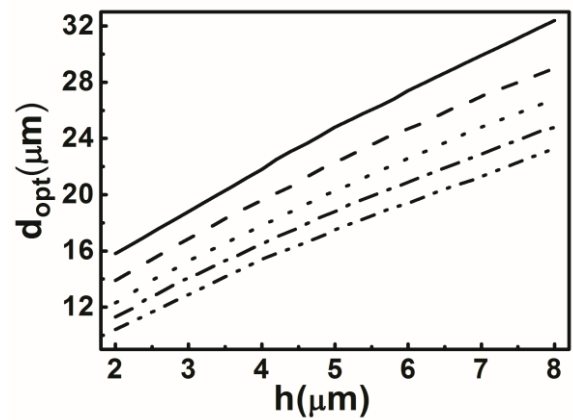

**Fig. 7.** Optimal NWSD $d_{opt}$ versus $h$ with $F_M = 40.0$ V/μm (solid), 45.0 V/μm (dash), 50.0 V/μm (dot), 55.0 V/μm (dash-dot), 60.0 V/μm (dash-dot-dot). The nanowall width is 0.16 nm.

## 4. Conclusion

Cold field electron emission from metallic nanowall array is analytically investigated with conformal mapping method. The field enhancement factor and transmission coefficient are obtained. The optimal nanowall spacing distance $d_{opt}$ for the largest field emission current density is found to be a linear function of nanowall height and depends on the applied field. The expression of $d_{opt}$ is obtained by fitting the numerical results.

## Appendices

### A. Elliptic integrals

The incomplete elliptic integrals of the first, second and third kinds, $F(\varphi|m)$, $E(\varphi|m)$ and $\Pi(n,\varphi|m)$ respectively, are specified in terms of the elliptic parameters $n$, $m$ ($0 \leq m \leq 1$), and the amplitude $\varphi$ ($0 \leq \varphi \leq \pi/2$) by

$$F(\varphi|m) = \int_0^{\varphi} (1-m\sin^2\vartheta)^{-1/2} d\vartheta, \tag{A1}$$

$$E(\varphi|m) = \int_0^{\varphi} (1-m\sin^2\vartheta)^{1/2} d\vartheta, \tag{A2}$$

$$\Pi(n,\varphi|m) = \int_0^{\varphi} (1-n\sin^2\vartheta)^{-1}(1-m\sin^2\vartheta)^{-1/2} d\vartheta. \tag{A3}$$

At $\varphi = \pi/2$, Eqs. (A1), (A2) and (A3) become corresponding complete elliptic integrals, i.e., $K(m) = F(\pi/2|m)$, $E_1(m) = E(\pi/2|m)$ and $\Pi_1(n,m) = \Pi(n,\pi/2|m)$.

### B. Barrier width

The potential barrier width along the emission direction satisfies

$$U(\Delta_y) - E_F = 0, \tag{B1}$$

where the potential energy $U(Z) = W_0 - e_0 V(0, h+Z)$, and $W_0$ the top of the tunneling barrier. Thus

$$V(0, h+\Delta_y) = \frac{W_0 - E_F}{e_0} = \frac{\phi}{e_0}. \tag{B2}$$

Substitution of Eqs. (9) and (10) into (B2) yields

$$\eta_\Delta = \frac{\phi B(0,\sigma')}{e_0 F_M r}. \tag{B3}$$

From Eqs. (11), (12), (4), (5) and (7),

$$h + \Delta_y = h + r \cdot \frac{\sigma'}{1-\sigma'} \cdot \frac{\Pi(b', \varphi_{\xi'_\Delta} | m')}{\Pi_1(\sigma', m) - K(m)}, \tag{B4}$$

where $\xi'_\Delta = -\sigma'' \sinh^2(\eta_\Delta/2) = -\sigma'' \sinh^2[\phi B(0,\sigma')/(2e_0 F_M r)]$, and

$$\varphi_{\xi'_\Delta} = \arcsin \sqrt{\frac{\sigma'' - \sigma'}{\sigma'' + \sigma' \sinh^{-2}[\phi B(0,\sigma')/(2e_0 F_M r)]}}. \tag{B5}$$

Let $\varphi_\Delta = \varphi_{\xi'_\Delta}$, then one obtains Eq. (20).

## C. Transmission coefficient and current density

Via JWKB approximation, the forward emission transmission coefficient is

$$D_y = \exp\left[-g_e \int_0^{\Delta_y} \sqrt{\phi - e_0 V(0, h+Z)} dZ\right], \tag{C1}$$

Let $t = e_0 V(0, h+Z)/\phi$ and $\eta_Z = \phi B(0,\sigma') t/(e_0 F_M r)$, then

$$\int_0^{\Delta_y} \sqrt{\phi - e_0 V(0, h+Z)} dZ = \frac{\phi^{3/2} B(0,\sigma')}{e_0 F_M r} \int_0^1 \sqrt{1-t} \frac{dZ}{d\eta_Z} dt, \tag{C2}$$

where, similarly to Eq. (B4), $Z = r \cdot \sigma'/(1-\sigma') \cdot \Pi(b', \varphi_{\xi'_Z} | m')/[\Pi_1(\sigma', m) - K(m)]$, $\xi'_Z = -\sigma'' \sinh^2(\eta_Z/2)$ and $\varphi_{\xi'_Z} = \arcsin(\sqrt{-\xi'_Z/[b'(-\xi'_Z + \sigma')]})$, so

$$\frac{dZ}{d\eta_Z} = \frac{r}{B(0,\sigma')} \sqrt{\frac{\sigma'' \sinh^2[\phi B(0,\sigma') t/(2e_0 F_M r)] + \sigma'}{\sigma'' \sinh^2[\phi B(0,\sigma') t/(2e_0 F_M r)] + 1}}. \tag{C3}$$

With Eqs. (C1), (C2) and (C3), one has Eq. (21).

According to the equation (4.13) of Ref. [9] the area emission current density at zero temperature is

$$J_A^{nws}(F_M) = z^{nw} \cdot (2d)^{-1} \cdot d_1^{3/2} \exp[-G_1], \tag{C4}$$

where $z^{nw} = 2e_0 \sqrt{2\pi m_e}/h_P^2$ and

$$G_1 = g_e \int_0^{\Delta_y} \sqrt{\phi - e_0 V(0, h+Z)} \, dZ, \tag{C5}$$

$$d_1^{-1} = \frac{g_e}{2} \int_0^{\Delta_y} \frac{1}{\sqrt{\phi - e_0 V(0, h+Z)}} \, dZ, \tag{C6}$$

Using Eqs. (C2) and (C3), one has

$$G_1 = \frac{g_e \phi^{3/2} h_D}{e_0 F_M}, \tag{C7}$$

$$d_1^{-1} = \frac{g_e \phi^{1/2}}{2 e_0 F_M} h_C, \tag{C8}$$

where $h_C$ is defined as Eq. (24). From (C4), (C7) and (C8) one can obtain

$$J_A^{nws}(F_M) = z^{nw} \left(\frac{2e_0}{g_e}\right)^{3/2} \cdot (2d)^{-1} \cdot \left(\frac{F_M}{\phi^{1/2} h_C}\right)^{3/2} \exp\left(-\frac{g_e \phi^{3/2} h_D}{e_0 F_M}\right). \tag{C9}$$

By denoting $z^{nws} = z^{nw} (2e_0/g_e)^{3/2} = e_0^{5/2}/(2^{3/4} \pi h_P^{1/2} m_e^{1/4})$, one obtains Eq. (23).